\documentclass[twocolumn]{aastex62}
\usepackage{natbib}
\usepackage{amsmath}
\usepackage{gensymb}

\newcommand{\kms}         {km\,s$^{-1}$}
\newcommand{\masyr}        {mas\,yr$^{-1}$}

\newcommand{\cravel}    {87.4}
\newcommand{\cravelu} {0.6} 
\newcommand{\cravell} {0.6}

\newcommand{\craveldisp}  {2.7}
\newcommand{\craveldispu} {0.5}
\newcommand{\craveldispl} {0.4}

\newcommand{\craml} {47} 
\newcommand{\cramlupper} {17}
\newcommand{\cramllower} {13}

\newcommand{\cramet}		{$-$1.95}
\newcommand{\crametlower} {0.05}
\newcommand{\crametupper} {0.06}

\newcommand{\crametdisp} 	  {0.18}
\newcommand{\crametdispupper}   {0.06}
\newcommand{\crametdisplower}    {0.08}

\shorttitle{Dynamical Histories of Crater~II and Hercules.}
\shortauthors{Fu et al.}

\begin{document}

\title{Dynamical Histories of the Crater~II and Hercules Dwarf Galaxies}

\author{Sal Wanying Fu}
\affiliation{Department of Physics and Astronomy, Pomona College, Claremont, CA 91711}

\author{Joshua D. Simon}
\affiliation{Observatories of the Carnegie Institution for Science, 813 Santa Barbara Street, Pasadena, CA 91101, USA}

\author{Alex G. Alarc\'{o}n Jara}
\affiliation{Departamento de Astronom\'{i}a, Universidad de Concepci\'{o}n, Casilla 160-C, 3349001 Concepci\'{o}n, Chile}

\correspondingauthor{Sal Wanying Fu}
\email{wanying.fu@pomona.edu}

\begin{abstract}

We investigate the possibility that the dwarf galaxies Crater II and Hercules have previously been tidally stripped by the Milky Way. We present Magellan/IMACS spectra of candidate member stars in both objects. We identify 37 members of Crater~II, 25 of which have velocity measurements in the literature, and we classify 3 stars within that subset as possible binaries. We find that including or removing these binary candidates does not change the derived velocity dispersion of Crater~II. Excluding the binary candidates, we measure a velocity dispersion of $\sigma_{V_{los}} = {\craveldisp}^{+\craveldispu}_{-\craveldispl}$~\kms, corresponding to $M/L =  \craml^{+\cramlupper}_{-\cramllower}$~M$_{\odot}/$L$_{\odot}$. We measure a mean metallicity of  [Fe/H] = \cramet$^{+\crametupper}_{-\crametlower}$, with a dispersion of $\sigma_{\mbox{[Fe/H]}} = \crametdisp^{+\crametdispupper}_{-\crametdisplower}$. Our velocity dispersion and metallicity measurements agree with previous measurements for Crater~II, and confirm that the galaxy resides in a kinematically cold dark matter halo. We also search for spectroscopic members stripped from Hercules in the possible extratidal stellar overdensities surrounding the dwarf. For both galaxies, we calculate proper motions using \textit{Gaia} DR2 astrometry, and use their full 6D phase space information to evaluate the probability that their orbits approach sufficiently close to the Milky Way to experience tidal stripping. Given the available kinematic data, we find a probability of $\sim40$\% that Hercules has suffered tidal stripping. The proper motion of Crater~II makes it almost certain to be stripped. 

\end{abstract}
9
\keywords{galaxies: dwarf --- galaxies: individual (Crater~II, Hercules) --- galaxies: kinematics and dynamics --- Galaxy: halo --- Local Group}

\section{Introduction}

\par The standard $\Lambda$CDM cosmological model predicts the existence of large numbers of dark matter subhalos surrounding Milky Way-like galaxies. The Milky Way's satellite dwarf galaxies, which are dark matter-dominated systems, are the luminous counterparts to some of the dark matter subhalos predicted in $\Lambda$CDM (e.g., \citealt{benson02},
\citealt{wetzel2016dsph}, \citealt{bullock2017}, \citealt{kim2017}). Representing galaxy formation on the smallest scales, dwarf galaxies are promising sites for understanding structure formation in the $\Lambda$CDM cosmology at the subhalo level. Studying their dynamics can also constrain the mass of their Milky Way halo host (e.g., \citealt{besla2007mcs}, \citealt{boylankolchin2013}, \citealt{barber2014}, \citealt{dierickx2017}, \citealt{patel2018}, \citealt{simon2018}, \citealt{eadie2018}, \citealt{watkins2018}). 

\par In recent years, the advent of wide-field photometric surveys has rapidly expanded the census of dwarf galaxies around the Milky Way (e.g., \citealt{belokurov2007sdss}, \citealt{bechtol2015des}, \citealt{koposov2015}, \citealt{drlicawagner2015des}, \citealt{laevens2015satellites}). These surveys, along with spectroscopic and deeper photometric follow up, have discovered several structurally peculiar satellites. Tidal interactions are frequently invoked in order to explain the properties of these systems. Hercules and Crater~II are two such dwarf galaxies. 

\par The ultra-faint dwarf galaxy Hercules was found in SDSS data in 2007 by \citet{belokurov2007sdss}. Since its discovery, deep photometric followup studies have uncovered substructures with stellar populations similar to that of Hercules beyond the tidal radius of the satellite (\citealt{coleman2007herculesstructure}, \citealt{sand2009herculesstructure}, \citealt{musella2012herc}, \citealt{deason2012stretch}, \citealt{roderick2015structure}, \citealt{garling2018rrlyr}). \citet{aden2009herculesmass} also presented tentative evidence for a velocity gradient across Hercules. The combination of these results, as well as the elongated shape of the galaxy, culminated in the hypothesis that Hercules is undergoing tidal disruption (e.g., \citealt{martin2010herculesdisrupt}, \citealt{blana2015orb}, \citealt{kuepper2017orbit}). 

\par With a half-light radius of $~$1100 pc, Crater~II is the fifth largest Milky Way satellite in physical extent \citep[][henceforth T16]{torrealba2016craterii}, trailing behind only the Magellanic Clouds, the Sagittarius dwarf spheroidal galaxy, and the newly discovered Antlia II. At a surface brightness of 31 mag arcsec$^{-2}$, Crater~II is also one of the most diffuse galaxies known. \citet[][henceforth C17]{caldwell2017craterIIspec} measured a line-of-sight velocity dispersion of $\sigma_{vlos} = 2.7 \pm 0.3$~\kms~for Crater~II, which is one of the coldest velocity dispersions resolved for any galaxy. While this velocity dispersion still renders Crater~II a dark matter-dominated system, with $M/L = 53^{+15}_{-11}$~M$_{\odot}/$L$_{\odot}$, Crater~II contains less dark matter within its half-light radius than other dwarfs with similar luminosities. The structural and kinematic properties of Crater~II are consistent with predictions from MOND \citep{mcgaugh2016mond}, but tidal stripping is necessary to explain the peculiar properties of Crater~II within $\Lambda$CDM (\citealt{sanders2018crater}, \citealt{fattahi2018tidalstripping}). 

\par Since both Hercules ($d=132$~kpc; \citealt{musella2012herc}) and Crater~II ($d=116$~kpc; \citealt{torrealba2016craterii}) lie far from the Milky Way center, it is not immediately obvious that they could have experienced significant tidal interactions with the Milky Way. The second data release \citep[DR2; ][]{gaiadr2} from the Gaia satellite \citep{gaiamission} provides strong proper motion constraints for many known Milky Way satellites (\citealt{gaia2018dsph}, \citealt{simon2018}, \citealt{fritz2018gaia}, \citealt{kallivayalil2018gaiadr2dsph}, \citealt{massari2018gaiaUFD}, \citealt{pace2018gaia}), which supplement the existing line-of-sight velocity data. The availability of full 6D phase space information for Milky Way satellites opens the possibility of detailed studies of their orbital properties.

\par The purpose of this work is to determine whether Hercules and Crater~II have previously undergone tidal interactions with the Milky Way. For the spectroscopic component of this work, we present Magellan/IMACS spectra of Crater~II members within its central 15$\arcmin$ in order to confirm its spectroscopic properties, build a larger member sample, and determine whether binary stars affect its observed velocity dispersion. We also target possible Hercules members in extratidal overdensities around the body of the dwarf to attempt to confirm their association with Hercules. We then use the 6D phase space information for each dwarf to evaluate the likelihood that they have made sufficiently close approaches to the Galactic Center to experience strong tidal effects. 

\par In Section \ref{sec:spectra}, we describe the instrument configuration, target selection, observations, and data reduction process for our spectroscopy. In Section \ref{sec:chemodynamics}, we describe our measurement and member selection procedures for Crater~II, and determine the velocity and metallicity dispersion of the galaxy. In Section \ref{sec:hercmem}, we briefly describe our search for kinematic members of Hercules beyond the tidal radius of the dwarf. In Section \ref{sec:tidaldisrupt}, we derive the orbits of Crater~II and Hercules, generate the probability distribution of each satellite's pericenter distances, and use tidal evolution tracks to infer the structural properties of their pre-stripping progenitors. In Section \ref{sec:discussion}, we discuss the implications of our results. In Section \ref{sec:conclusion}, we summarize our conclusions and offer some final remarks. 

\section{Spectra Acquisition and Reduction}
\label{sec:spectra}
\subsection{Spectrograph Set-up and Observation Overview}

\par We make our observations using the IMACS spectrograph \citep{imacs} on the Magellan Baade telescope. We use the $f$/4 camera and the $1200~\ell$~mm$^{-1}$ grating to provide R~$\sim$~11000 spectra covering the Ca triplet lines in the near-infrared. Specific details of the spectrograph function and set-up are provided in \citet{tucIII} and \citet{li2017eri2}.

\par A typical observing procedure for this study is to acquire two science exposures lasting $\sim$~40 minutes each. After every set of science frames, we take calibration frames using comparison arclamps and flatfield lamps at the same pointing position. For our arc frames, we use He, Ne, Kr and Ar lamps. 

\subsection{Crater~II Target Selection and Observation} 

\begin{deluxetable*}{cccccccccc}
\tablecaption{Observations of Crater II and Hercules}
\tablehead{\colhead{Object} & \colhead{Mask Name} & \colhead{$\alpha$ (J2000)} &
\colhead{$\delta$ (J2000)} & \colhead{Slit PA} & 
\colhead{$t_{exp}$} &  \colhead{MJD of Observation} & 
\colhead{\# of Slits} & \colhead{Seeing} &  \colhead{S/N ($i=$18)} \\
\colhead{} & \colhead{} & \colhead{(h:m:s)} & 
\colhead{(d:m:s)} & \colhead{(deg)} &
\colhead{(s)} & \colhead{} & \colhead{} &
\colhead{(arcsec)} & \colhead{(pixel$^{-1}$)}}
\startdata
Crater II & Mask 1        & 11:49:44.0 & $-$18:27:00 & 65  & 9600  & 58201.3 & 78 &    $\sim$1.0 & 35 \\
Crater II & Mask 2        & 11:48:22.4 & $-$18:28:57 & 1   & 9000  & 58202.3 & 80 &    $\sim$0.6 & 45 \\
Crater II & Mask 3        & 11:49:05.0 & $-$18:33:00 & 330 & 6500  & 58203.3 & 76 &    $\sim$0.6 & 35 \\
Crater II & Mask 4        & 11:49:14.0 & $-$18:17:15 & 23  & 1800  & 58203.3 & 44 &    $\sim$0.6 & 20 \\
Hercules   & OD13.2, Mask 1 & 16:29:56.9 & $+$12:54:30 & 182 & 4800  & 57567.0 & 57 &    $\sim$1.0 & 20 \\
Hercules   & OD13.2, Mask 2 & 16:29:56.9 & $+$12:54:30 & 178 & 1686  & 57926.1 & 59 &    $\sim$0.7 &  5 \\
Hercules   & OD16         & 16:28:41.0 & $+$12:53:17 & 183 & 10800 & 57924.2 & 82 &    $\sim$1.2 & 25 \\
\label{tab:observinglog}
\enddata
\end{deluxetable*}

\par
\begin{figure*}
\epsscale{1.2}
\plotone{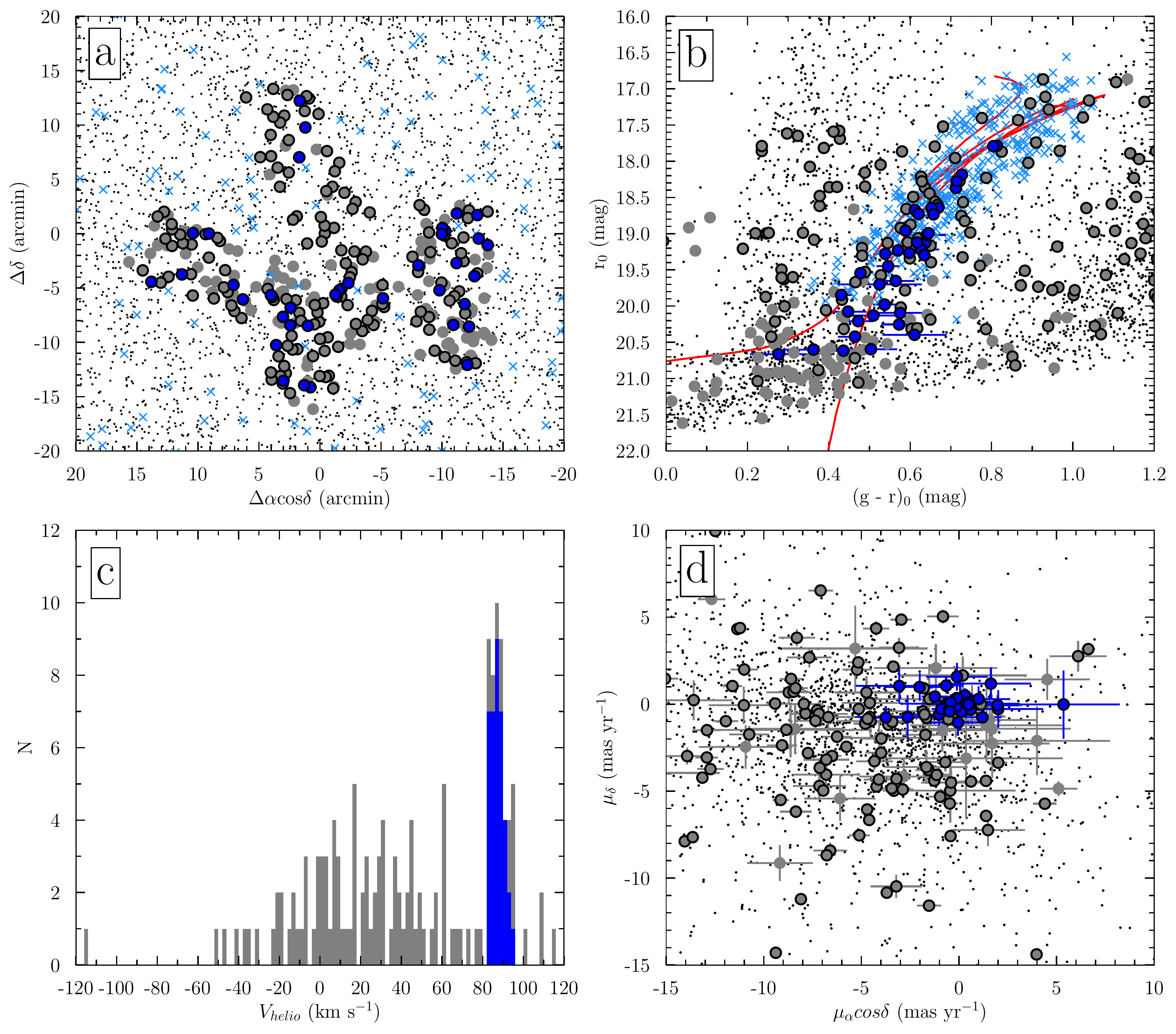}
\caption{Observation targets and selection of Crater~II members. Gray circles designate stars that were observed; those outlined in black are stars for which we obtained usable spectra. Blue circles designate stars that we determine to be photometric, spectroscopic and astrometric members of Crater~II. Light blue crosses (in the upper panels) are all of the stars that were observed by C17. a) On-sky distribution of stars within our survey area. The small dots are all of the stars within 20\arcmin\ of the center of Crater~II. b) Dereddened PS1 photometry of the stars within the same area shown in a). The overplotted Padova isochrone \citep{padova}, shifted to the distance of Crater~II, corresponds to a stellar population of ${\rm [Fe/H]} = -2.00$ and age = 12 Gyr. We were unable to obtain a velocity measurement for the bright star at $r_0 = 16.87$~mag, $(g-r)_0 = 1.13$~mag because its spectrum landed in a chip gap on the detector mosaic. c) Distribution of heliocentric velocities for our targets; bins are 2 \kms\ wide. The gray histogram corresponds to the velocity distribution of all the stars in our survey, while the blue histogram is the velocity distribution of confirmed Crater~II members. The velocity signature of Crater~II, centered at $\sim$87 \kms, is clearly visible. d) Proper motion distribution of Crater~II members and candidates within the region plotted in panel a). The proper motion of the dwarf deviates from that of the background.}
\label{fig:cramems}
\end{figure*}

 We selected spectroscopic targets in Crater~II using Pan-STARRS photometry \citep[PS1; ][]{panstarrs}, and corrected for extinction using the dust maps of \citet{sfd1998} and the extinction coefficients of \citet{schlafly2011}. We used a Padova isochrone \citep{padova} corresponding to ${\rm [Fe/H]} = -2.0$ and age = 12~Gyr as well as the sample of spectroscopic members from C17 to guide our selection of likely Crater~II members. In addition to the known members, we chose candidate red giant branch (RGB) stars within 0.12 mag of the Padova isochrone toward redder colors and within 0.18 mag of the isochrone on the blue side. We chose mask positions and orientations to maximize the number of C17 stars observed. Given the size of the IMACS $f/4$ field of view, our targets all lie within $\sim15\arcmin$ of the center of Crater~II (see Figure \ref{fig:cramems}), making our survey area smaller than that of C17. 
\par We observed a total of four slitmasks targeting candidate member stars in Crater~II on three nights in 2018 March. Observing conditions were clear on all three nights, and seeing was typically below 1$\arcsec$. Table \ref{tab:observinglog} presents the overview of observations for the Crater~II masks. 

\subsection{Hercules Target Selection and Observation}

\par We observed a total of three slitmasks targeting candidate member stars in the extratidal densities surrounding Hercules. Using Sloan Digital Sky Survey photometry \citep[SDSS-III; ][]{eisenstein2011sdss}, dereddened with the dust maps of \citet{sfd1998} and extinction coefficients of \citet{schlafly2011}, we selected candidate Hercules RGB members based on their proximity to the fiducial sequence of M92 \citep{clem08}. Our observations targeted the overdensities (ODs) 13.2 and 16 as designated in \citet{roderick2015structure}, which are the most significant overdensities surrounding the galaxy that the authors identified (also see \citealt{sand2009herculesstructure}).

\par We observed Hercules during two nights in 2016 June and one night in 2017 June. Overall, we observed two masks in OD13.2 and one mask in OD16. Spectroscopy of the second mask targeting OD13.2 is quite shallow because observations were cut short by high winds. Table \ref{tab:observinglog} presents the overview of our Hercules observations.

\subsection{Data Reduction}
\label{sec:datreduc}

\par We began our data reduction process by using the COSMOS software\footnote{\href{http://code.obs.carnegiescience.edu/cosmos}{http://code.obs.carnegiescience.edu/cosmos}} (\citealt{dressler2011cosmos}, \citealt{cosmos}) to derive approximate wavelength solutions. The two-dimensional map of each slit mask produced by COSMOS was then used as the starting point for reductions with a modified version of the DEEP2 pipeline originally developed for Keck/DEIMOS (\citealt{cooper2012deeppipeline}, \citealt{newman2013deeppipeline}). Further details on the data reduction process and modifications to the DEEP2 pipeline can be found in \citet{tucIII}, \citet{simongeha2007} and references therein. 

\par We reduced every set of science exposures using the corresponding set of calibration frames. For masks with multiple sets of exposures taken on the same night, we combined the extracted 1D spectra using inverse-variance weighting. 

\section{Chemodynamics of Crater~II members}
\label{sec:chemodynamics}

\subsection{Radial Velocity Measurements}

\par We measured radial velocities using the procedures described by \citet{simongeha2007}, \citet{tucIII}, \citet{li2017eri2} and associated papers. Using the same IMACS configuration described in Section \ref{sec:spectra}, we observed a set of bright, metal-poor stars to serve as the radial velocity template spectra. We also observed the hot, rapidly rotating star HR~4781 to use as a telluric template for measuring A-band velocity corrections. The details of our template observations can be found in \citet{tucIII} and references therein. 

\par We measured the radial velocity of each science spectrum by minimizing its $\chi^2$ fit to the template spectrum (\citealt{simongeha2007}, \citealt{newman2013deeppipeline}). We use the cool, metal-poor red giant HD~122563 as our template for the science spectra. 
We use the telluric template to fit the A-band absorption of every science spectrum. The measured velocity of the A-band corrects for any mis-centering of each star in its slit. These corrections are generally less than 6~\kms, and show a systematic dependence on the position of the slit on the mask in the direction parallel to the slits \citep{li2017eri2}. We model this dependence as a quadratic function, and apply the modeled A-band correction for stars with poor A-band measurements.

\par Per procedure in \citet{simongeha2007}, we calculate the statistical uncertainty on each velocity measurement by performing Monte Carlo simulations, in which we add randomly distributed noise to the spectrum and redo the template fitting. We define the uncertainty as the standard deviation of the Monte Carlo measurements after removing $>5\sigma$ outliers from the distribution. We add the Monte Carlo uncertainty in the velocity measurements, the Monte Carlo uncertainty in the A-band corrections, and the 1.0~\kms\ systematic uncertainty determined by \citet{tucIII} in quadrature to obtain the total uncertainty on each radial velocity measurement. 

\subsection{Metallicity Measurements}

\par We measured metallicities for stars in Crater~II by using the five-parameter calcium triplet (CaT) calibration of \citet{carrera2013cat}, which requires the equivalent widths of the CaT lines. Following the procedures in \citet{simon2015}, we fit each of the CaT lines with the sum of a Gaussian and Lorentzian profile. We determined the equivalent widths of each line by integrating under the fitted profiles, and use the summed equivalent widths of all three lines for the \citet{carrera2013cat} absolute $V$-magnitude calibration. We calculated statistical uncertainties on the equivalent widths using the uncertainties on the Gaussian and Lorentzian integrals. Per \citet{tucIII}, the systematic uncertainty on the summed equivalent widths of the CaT lines is 0.32~\AA. To obtain the total measurement uncertainty, we added the statistical and systematic equivalent width uncertainties in quadrature. 

\subsection{Membership Determination for Crater II}
\label{sec:craIImem}

\par We present the spatial distribution, color-magnitude diagram, velocity distribution, and proper motion distribution of the observed stars in Figure \ref{fig:cramems}. From the sample of reliable velocity measurements, it is evident that Crater~II consists of stars in a narrow range of velocities around 87~\kms, consistent with the results of C17. We begin our Crater~II member determination process by selecting all the stars that fall within  3~$\sigma_{V_{los}}$ of the mean $V_{los}$ measurement from C17. The majority of the stars resulting from this selection have photometry consistent with Crater~II membership (Figure \ref{fig:cramems}b). We impose a final membership requirement that all stars must have \textit{Gaia} DR2 \citep{gaiadr2} proper motions that are consistent with the bulk motion of the dwarf. Since most stars that pass the velocity and photometric criteria have proper motions consistent with each other, we apply a final proper motion selection criterion in which all remaining stars must fall within 3 $\sigma$ of the overall proper motion of Crater~II.  

\par From this process, we identify 37 Crater~II members. Of these 37 stars, 25 overlap with likely Crater~II members in C17\footnote{Since the catalog accompanying C17 does not provide membership determinations, we identify Crater~II members from C17 using the same membership selection criteria that we use for our study.}, providing a time baseline of nearly 2 years for velocity changes due to binary orbital motion. Of the 25 overlap stars, we identify 3 that have velocity measurement differences close to or beyond 2 $\sigma$ as binary candidates. These stars are PSO~J114820.50$-$183233.3 ($\Delta V/V_{err} = 1.93$),  PSO~J114825.96$-$183223.5 ($\Delta V/V_{err} = 2.15$), and PSO~J114829.90$-$182402.2 ($\Delta V/V_{err} = 2.35$). The results of our membership selection are presented in Figure \ref{fig:cramems} and Table \ref{tab:cra2mem}. 

\subsection{Velocity and Metallicity Dispersion of Crater~II}
\label{sec:craIIchemodynamics}

\par We measure the mean velocity and velocity dispersion of the Crater~II member stars by using the Markov Chain Monte Carlo code \textit{emcee} \citep{emcee} to maximize the Gaussian likelihood function defined by \citet{walker06}. If we include the 3 likely binary stars identified in the previous section in our sample, we measure the bulk velocity of Crater~II to be 87.4$^{+0.5}_{-0.5}$~\kms, and the velocity dispersion to be $\sigma_{Vlos} = 2.7^{+0.4}_{-0.4}$~\kms. Without the binary candidates, we measure the bulk velocity of Crater~II to be $\cravel^{+\cravelu}_{-\cravell}$~\kms, and the dispersion to be $\sigma_{Vlos} = \craveldisp^{+\craveldispu}_{-\craveldispl}$ \kms. Thus, we find that the inclusion or exclusion of binary stars does not significantly affect the derived velocity dispersion of Crater~II. Since the presence of binary stars is a general concern when inferring dynamical masses of dwarf galaxies (e.g., \citealt{minor2018}, \citealt{spencer2018}), we adopt the velocity dispersion measured without the binaries and use that value for the remainder of the paper. Using the equation from \citet{wolf2010}, we calculate that the mass-to-light ratio within the half-light radius of Crater~II is $M/L$ = \craml$^{+\cramlupper}_{-\cramllower}$~M$_{\odot}/$L$_{\odot}$. Our velocity and dynamical mass measurements are consistent with those of C17, confirming that while Crater~II is still dark matter-dominated, it also resides in a kinematically cold dark matter halo.

\par We measure the mean metallicity of Crater~II to be [Fe/H] = \cramet$^{+\crametupper}_{-\crametlower}$, with a corresponding dispersion of $\sigma_{\mbox{[Fe/H]}} = \crametdisp^{+\crametdispupper}_{-\crametdisplower}$~dex. While our metallicity dispersion measurement is consistent within the uncertainties with that of C17, we are only able to resolve a metallicity dispersion at the 98\% ($<3 \sigma$) confidence level. However, it is possible that the galaxy contains a metallicity gradient, and that we are measuring a slightly smaller metallicity dispersion because our survey area was smaller than that of C17. 

\begin{figure*}
\epsscale{1.2}
\plotone{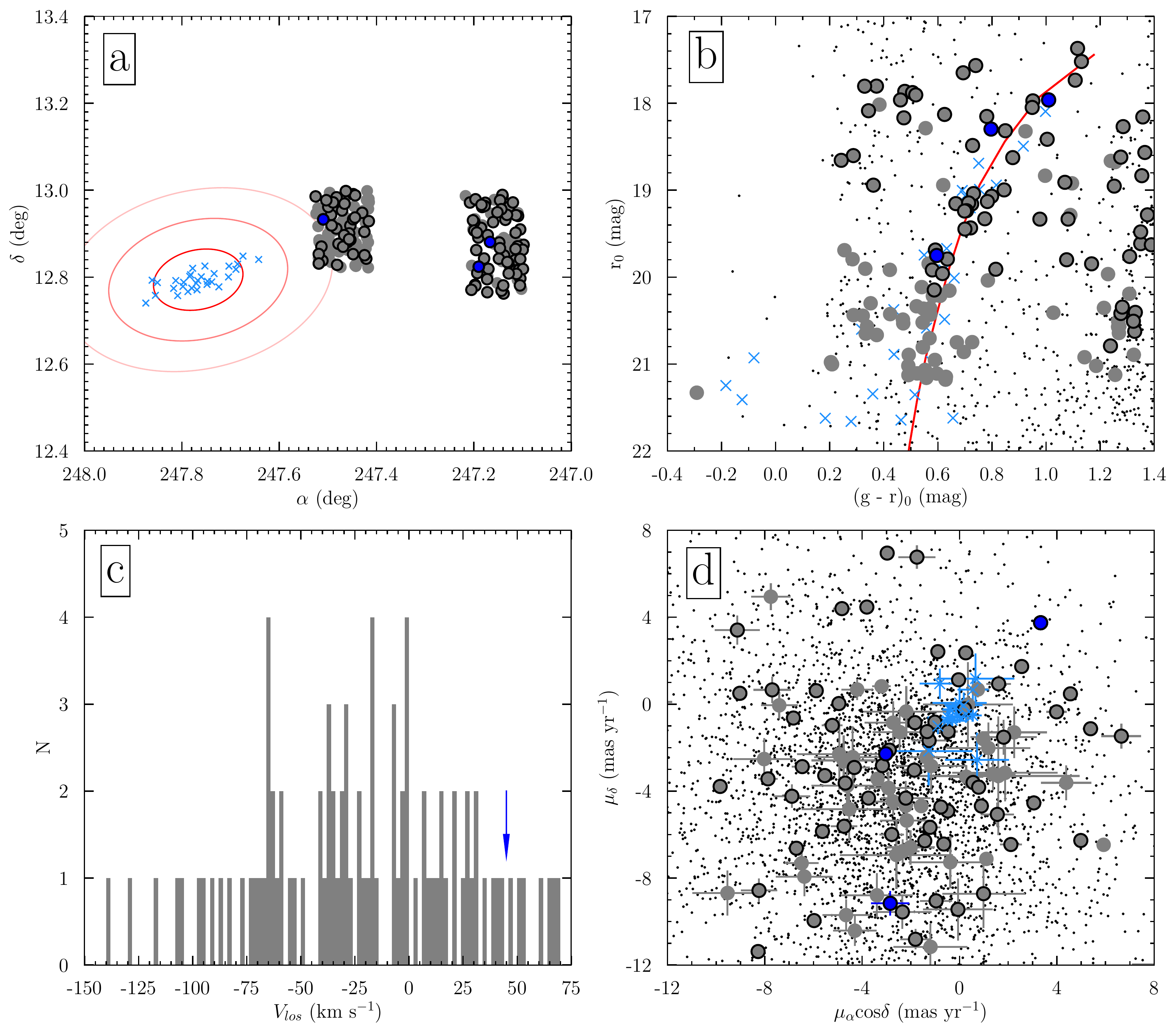}
\caption{Observation targets and selection of candidate Hercules members in the dwarf's extratidal overdensities. Gray circles designate stars that were observed; those outlined in black are stars for which we obtained usable spectra. Blue circles designate stars whose line of sight velocity and photometry are consistent with Hercules membership, but whose proper motions suggest that they are Milky Way foreground stars. Light blue points in panels a), c), and d) are the Hercules members identified by \citet{simongeha2007}. a) On-sky distribution of stars that we observed. The patch between $\alpha=247.6$\degree~and $\alpha=247.4$\degree~is OD13.2 as designated by \citet{roderick2015structure}. The patch between $\alpha=247.2$\degree~and $\alpha=247.0$\degree~is OD16 from the same study. The red lines show the $r_h$, $2r_h$, and $3r_h$ ellipses of Hercules. b) Dereddened SDSS photometry of the stars in the areas of OD13.2 and OD16. We overplot the fiducial sequence of M92 from \citet{clem08}, shifted to a distance of 132~kpc. c) Histogram of velocities measured from stars in our sample. Bins are 2~\kms~wide. The blue arrow corresponds to the line of sight velocity of Hercules. d) Proper motion distribution of Hercules members from SG07, and of candidate members from our study. The three stars whose velocity and photometry are consistent with Hercules membership have proper motions discrepant from that of the body of the dwarf.}
\label{fig:hercmems}
\end{figure*}

\section{Membership Determination for Hercules}
\label{sec:hercmem}
\par From our spectroscopic data set, we aim to identify kinematic members of Hercules in the possible extratidal overdensities around the dwarf. We present the results of that exercise in Figure  \ref{fig:hercmems}. We begin by selecting all stars that fall within 3~$\sigma$ of the line-of-sight velocity of Hercules according to the values from \citet{simongeha2007}. We then select stars whose photometric properties are consistent with being Hercules members. From these criteria, we identify 3 member candidates: one in OD13.2 and two in OD16. However, these stars have proper motions that are inconsistent with the bulk motion of the dwarf by well over 3 $\sigma$, and must therefore be foreground Milky Way stars (see Figure \ref{fig:hercmems}d). Thus, our data set does not include any RGB stars brighter than $r \sim 20$ that are kinematically associated with Hercules. Given the limited depth of our spectroscopy, we cannot rule out the presence of fainter RGB stars or main sequence stars associated with Hercules in these overdensities. 

\section{Tidal Interaction Scenarios over Milky Way Parameter Space}
\label{sec:tidaldisrupt}

\begin{deluxetable}{cc}
\tablecaption{6D Parameters and Derived Orbital Parameters for Crater~II}
\tablehead{\colhead{Parameter} & \colhead{Value}}
\startdata
$\alpha$ & 177.3 deg \\
$\delta$ & $-$18.4 deg \\
$D_{\odot}$ & $117.5 \pm 1.1$ kpc \\
$\mu_{\alpha}\cos{\delta}$ & $-0.17 \pm 0.07$ \masyr \\
$\mu_{\delta}$ & $-0.07 \pm 0.05$ \masyr \\
$V_{\mbox los}$ & $\cravel^{+\cravelu}_{-\cravell}$~\kms  \\
\hline 
$r_{peri}$ & 37.7$^{+18.0}_{-13.3}$~kpc \\
Orbital Period & 2.2$^{+0.7}_{-0.4}$ Gyr \\
Eccentricity & 0.56$^{+0.12}_{-0.11}$ \\
\enddata
\tablecomments{6D parameters for Crater~II used for the kinematic analysis in Section \ref{sec:craIIorb}, and the orbital parameter summary statistics for the spherical halo case. The heliocentric distance measurement is from T16. Proper motion and line-of-sight velocity are from this work.}
\label{tab:craIIorb}
\end{deluxetable}

\begin{deluxetable}{cc}
\tablecaption{6D Parameters and Derived Orbital Parameters for Hercules}
\tablehead{\colhead{Parameter} & \colhead{Value}}
\startdata
$\alpha$ & 247.77 deg \\
$\delta$ & 12.79 deg \\
$D_{\odot}$ & 132 $\pm$ 6 kpc \\
$\mu_{\alpha}\cos{\delta}$ & $-0.16 \pm 0.09$ \masyr \\
$\mu_{\delta}$ & $-0.41 \pm 0.07$ \masyr \\
$V_{\mbox los}$ & $45.0 \pm 1.1$ \kms \\
\hline
$r_{peri}$ & 47.2$^{+27.0}_{-21.6}$ kpc \\
Orbital Period & 3.5$^{+2.0}_{-1.3}$ Gyr \\
Eccentricity & 0.69$^{+0.11}_{-0.08}$ \\
\enddata
\tablecomments{6D parameters for Hercules used for the kinematic analysis in Section \ref{sec:hercorb}, and the orbital parameter summary statistics for the spherical halo case. The heliocentric distance measurement is from \citet{musella2012herc}, the line-of-sight velocity measurement is from \citet{simongeha2007}, and the proper motion measurements are from this work.} 
\label{tab:hercorb}
\end{deluxetable}

\begin{figure}[!ht] 
\includegraphics[scale=0.50]{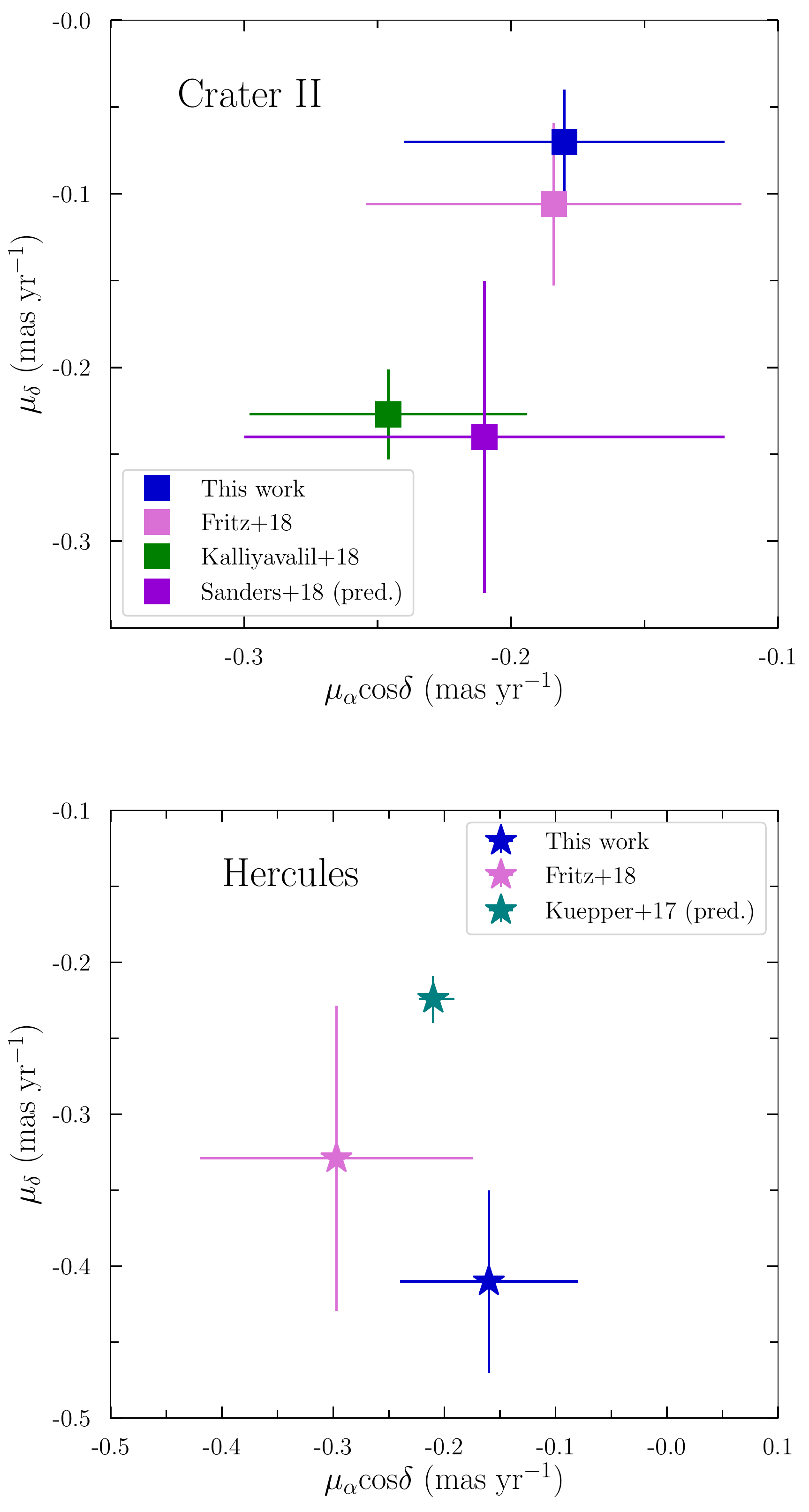}.
\caption{Comparison of the proper motion measurements for Hercules (star symbol) and Crater~II (square symbol) to existing measurements in the literature, as well as predictions from theoretical studies of the tidal disruption of each dwarf (\citealt{kuepper2017orbit}, \citealt{sanders2018crater}, \citealt{fritz2018gaia}, \citealt{kallivayalil2018gaiadr2dsph}). We find that our measured proper motions are generally consistent with literature measurements within the uncertainties.}
\label{fig:pmcomparison}
\end{figure}

\par We explore the orbital properties of Crater~II and Hercules using the open-source code \textit{galpy} \citep{bovy2015galpy}. We adopt the solar motions from \citet{schonrich2010solar} and set up our Milky Way potential using the results from \citet{mcmillan2017solar}. 

\par We calculate the proper motion of Crater~II by combining stars from our sample and those we identify as members from the Caldwell sample. We compare our proper motion measurement to those of \citet[][]{fritz2018gaia} and \citet[][]{kallivayalil2018gaiadr2dsph} in Figure \ref{fig:pmcomparison}, and find that our measurements are in good agreement with F18. While the \citet{kallivayalil2018gaiadr2dsph} measurement is consistent with ours in the RA direction, it deviates in the Dec direction by about 3~$\sigma$. Our measurement is also consistent with the proper motion predicted by \citet[][]{sanders2018crater} for the case of a tidally stripped Crater~II (Figure \ref{fig:pmcomparison}).

\par We calculate the proper motion of Hercules by taking the weighted mean proper motion of individual members from \citet{simongeha2007} and \citet{aden2009herculesmass}. Our measurement is consistent with the measurement from \citet{fritz2018gaia} to within 1~$\sigma$ (Figure \ref{fig:pmcomparison}), but with a smaller uncertainty given the larger sample. 

\par Tables \ref{tab:craIIorb} and \ref{tab:hercorb} present the observed properties that we use to initialize the orbits of the respective satellites. We obtain the final proper motion uncertainties by adding the weighted standard deviation of the mean and the \textit{Gaia} DR2 systematic floor of 0.035~\masyr\ \citep{gaia2018dsph} in quadrature. 

\par For both satellites, we integrate a fiducial orbit using point-estimates of the mass of the potential and their 6D phase space information. To determine the probability distribution for the pericenter distances of their orbits, we also run a Monte Carlo simulation where we model $M_{MW}$, $\mu_{\alpha}\cos{\delta}$, $\mu_{\delta}$ and $D_{\odot}$ as Gaussian distributions, with the width of each Gaussian set by the uncertainty of the corresponding parameter (see Tables \ref{tab:craIIorb} and \ref{tab:hercorb} and \citealt{mcmillan2017solar}). Since the position and line-of-sight velocity have negligible uncertainties in comparison to parameters such as distance and proper motion, we fix those values for our analysis. 

\par For each simulated set of parameters, we integrate the orbit and find its pericenter distance. We also conduct this exercise for flattened Milky Way halos with axis ratios $c/a$ ranging from 0.5 to 1.0 in steps of 0.1. For both galaxies, a more spherical halo results in larger pericenter distances, but the general conclusion of this study would not change by adopting different flattening values. For the sake of brevity, we discuss our results for the case of a spherical halo in the following sections. However, we make the code used for the analysis in this section available online, and include an illustration of the effects of incorporating flattening for the orbits of both dwarfs.\footnote{\href{https://github.com/swfu/DwarfTidalStripping}{https://github.com/swfu/DwarfTidalStripping}}

\subsection{Tidal Radius Calculation}
\label{sec:tidalradius}

\par A key question for this study is how small a satellite's pericenter distance must be in order for it to experience tidal effects from the Milky Way. For a satellite orbiting a larger host, the stars outside of the satellite's tidal radius will be lost to the host galaxy's tidal forces. To approximate the tidal radius of a satellite, we use the following equation for the Jacobian radius, $r_J$, referenced from \citet{bt08} and adapted for our parameters of interest:

\begin{equation}
    r_{tidal} \sim r_J = \left(\frac{m_{sat}}{3M(R)}\right)^{1/3}R,
\end{equation}

\noindent where $m_{sat}$ is the mass of the satellite, $R$ is the Galactocentric distance of the satellite, and $M(R)$ is the enclosed of mass of the Milky Way within that Galactocentric distance. Thus, the tidal radius of a satellite decreases with decreasing pericenter distance. We deem it likely for a satellite to have experienced tidal stripping if the pericenter of its orbit is smaller than the Galactocentric distance where its tidal radius is equal to 3 times its half-light radius; that is, where $r_{tidal}$ = $3r_h$. We choose this number because for a galaxy with a Plummer stellar profile, 90\% of its stars are located within three half-light radii. 

\par
\begin{figure}[!ht] 
\includegraphics[scale=0.55]{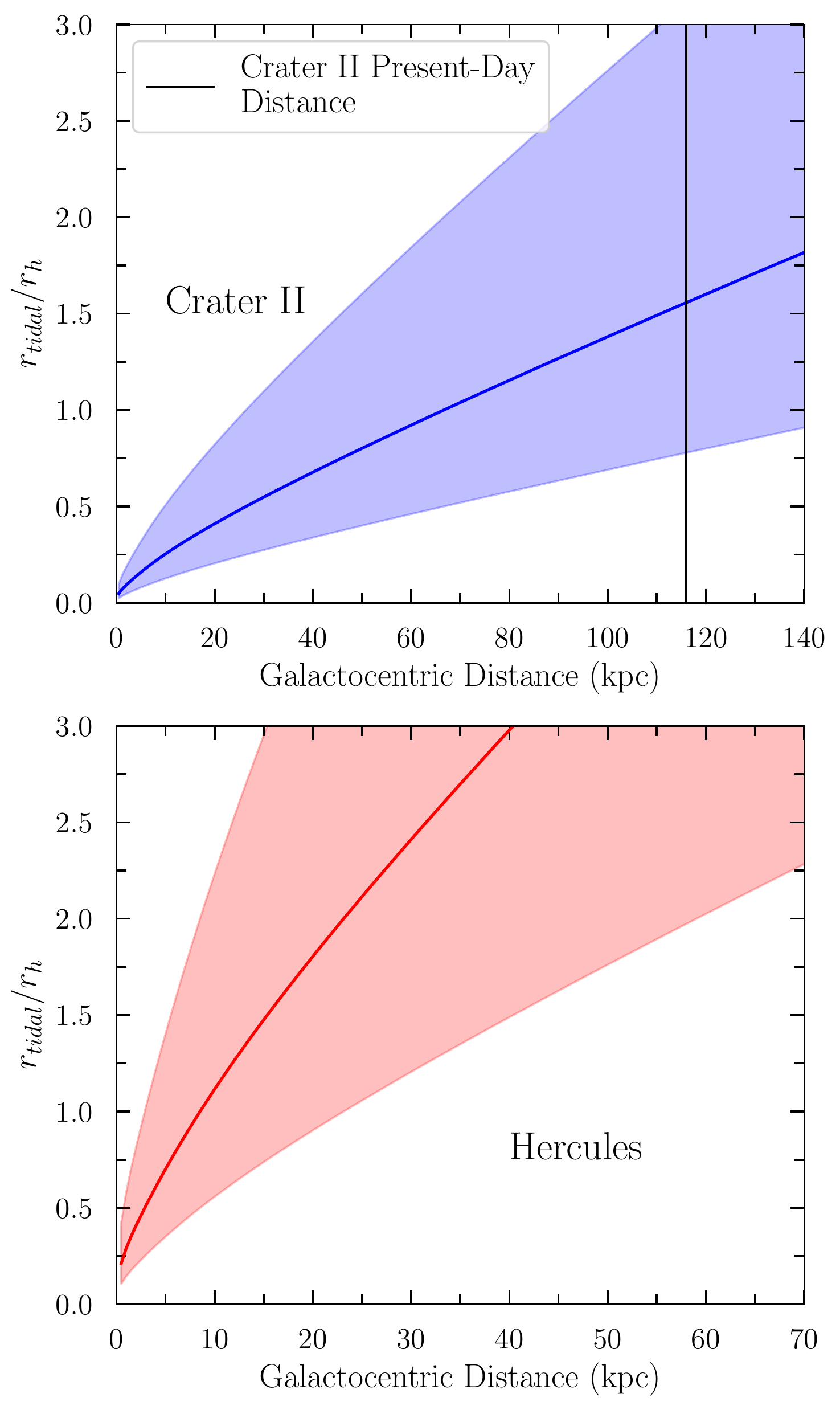}
\caption{Ratio of tidal radius to half-light radius for each satellite as a function of their Galactocentric distance. The shaded region encapsulates the region of uncertainty obtained by scaling the tidal radius equation up and down by a factor of 2. For Crater~II, a non-negligible fraction of its stars are vulnerable to being stripped even at its present-day position.}
\label{fig:tidalradius}
\end{figure}

 We use \textit{galpy} \citep{bovy2015galpy} to calculate $M(R)$, incorporating contributions from the bulge, the disk, and the halo, again employing the mass distribution from \citet{mcmillan2017solar}. For each satellite, we find $m_{sat}$ using the results of \citet{wolf2010} to calculate the mass of the satellite enclosed within its half-light radius. For Crater~II, we calculate an enclosed mass of $7.4 \times 10^6$~M$_{\odot}$. Using the velocity dispersion from \citet{simongeha2007} and the half-light radius from \citet{munoz2018imaging} for the Plummer profile, we calculate for Hercules an enclosed mass of $5.2 \times 10^6$~M$_{\odot}$. We present the results of these calculations in Figure \ref{fig:tidalradius}.

\par There are two major sources of uncertainty in this calculation: (1) Since we use only the mass of the satellite enclosed within the half-light radius, our calculation actually yields a lower limit on the tidal radius at any given Galactocentric distance. In fact, the tidal radius must be larger because we do not account for the extended mass of the dark matter halo in which the satellite is embedded. This, in turn, implies that the satellite's pericenter distance must actually be smaller than calculated in order for the satellite to experience tidal stripping. (2) The Jacobian radius is not a perfect approximation of the tidal radius. To account for these uncertainties, Figure \ref{fig:tidalradius} illustrates not only the relationship derived from Equation 1, but also a region of uncertainty, obtained by scaling Equation 1 up and down by a factor of 2. 

\par We find that for Crater~II, $r_{tidal}/r_h = 1.5$ at its present-day location ($R=116$~kpc). Thus, even at its current galactocentric distance Crater~II will suffer stripping unless it has retained a massive halo. For Hercules, we find that $r_{tidal}/r_h = 3$ at 40~kpc from the Galactic Center. Thus, Hercules must have an orbital pericenter smaller than 40~kpc for the satellite to be tidally stripped. 

\par The results of this analysis are also available on the Github link provided in the previous section. 

\subsection{Crater II}
\label{sec:craIIorb}

\par
\begin{figure*}
\epsscale{1.2}
\plotone{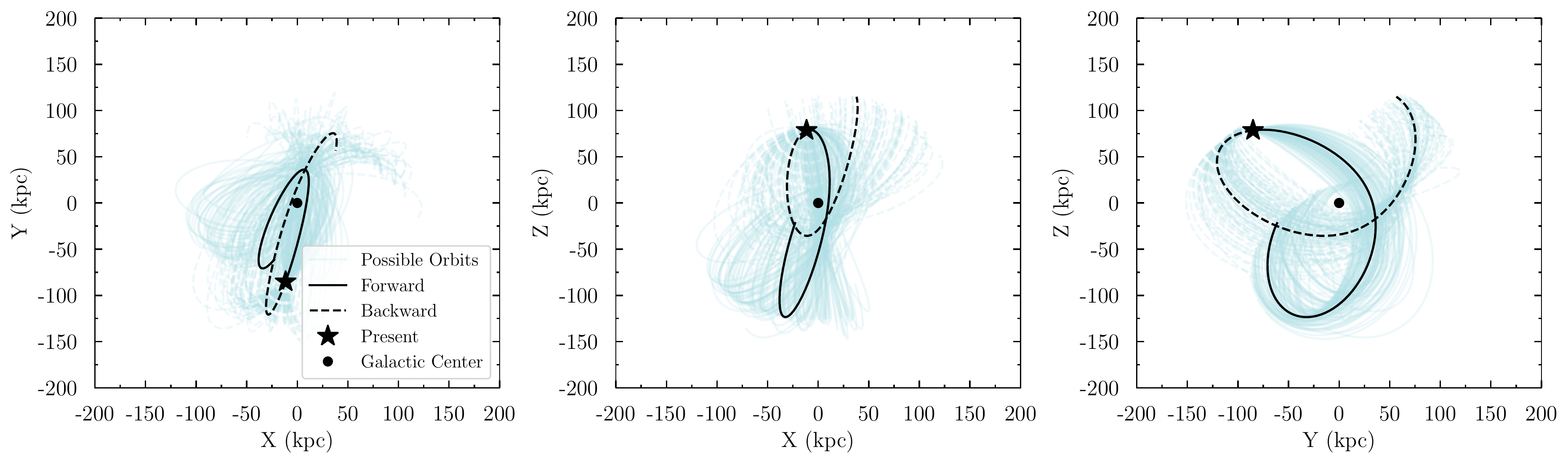}
\caption{Fiducial orbit of Crater II (black), integrated forward and backward for 2.5~Gyr. In this orbit, Crater~II passed pericenter 1.5~Gyr ago, approaching within 33~kpc of the Milky Way. Light-blue orbits correspond to other possible orbits within the proper motion uncertainties.}
\label{fig:craIIfid}
\end{figure*}

 In Figure \ref{fig:craIIfid}, we present the fiducial orbit of Crater~II. In this orbit, the pericenter is 33 kpc, and the orbital period is 2.1~Gyr. The last pericentric passage Crater~II made was 1.5 Gyr ago. It also recently passed apocenter, and is now on its way back toward the Milky Way.

\par
\begin{figure*}
\epsscale{1.2}
\plotone{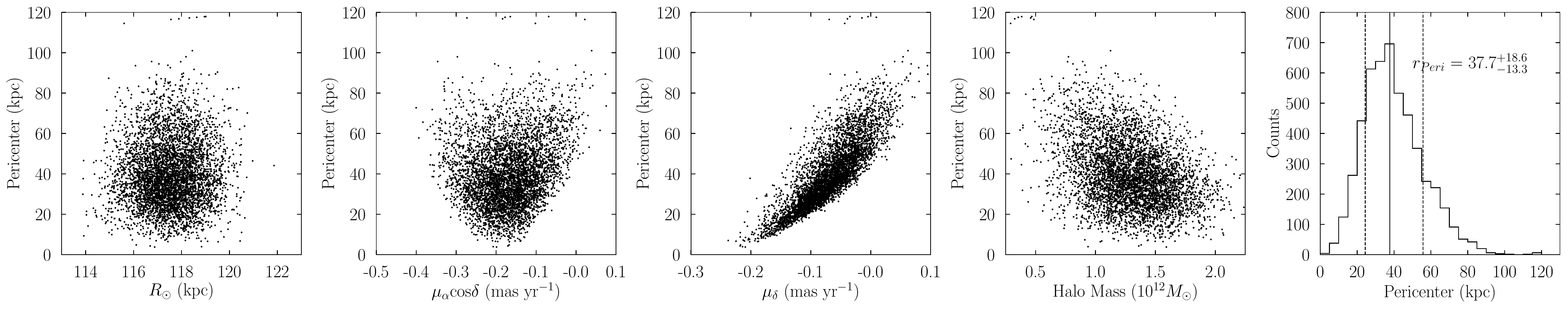}
\caption{Monte Carlo simulations for the orbital pericenter of Crater~II, integrated in a spherical potential. Increasingly negative $\mu_{\delta}$ values correspond to lower pericenter distances; thus currently existing proper motion measurements for Crater~II in the literature are consistent with the tidal disruption hypothesis. All possible orbits for Crater~II keep the dwarf within 120~kpc of the Galactic Center, lending credence to the idea that tidal effects may be responsible for its unusually small velocity dispersion and large size \citep{sanders2018crater}.}
\label{fig:craIIMC}
\end{figure*}

 In Figure \ref{fig:craIIMC}, we present the results of our Monte Carlo simulation for Crater~II. Although our $\mu_{\delta}$ measurement differs from that of \citet{kallivayalil2018gaiadr2dsph}, increasingly negative $\mu_{\delta}$ values correspond to smaller pericenter distances. Thus, current proper motion measurements for Crater~II in the literature also support tidal stripping scenarios.  

\subsection{Tidal Evolution of Crater~II}
\label{sec:craIIevo}


\begin{figure*}
\epsscale{1.2}
\plotone{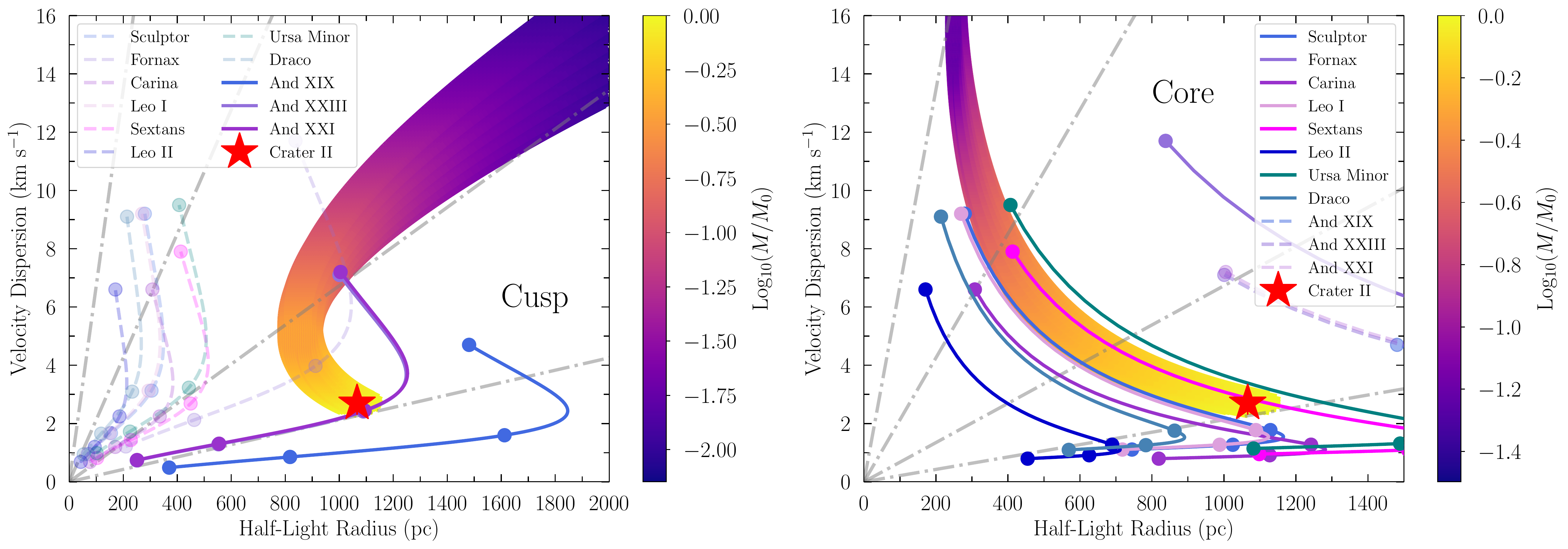}
\caption{Possible tidal evolution of Crater~II. We infer possible analogs to the progenitor of Crater~II by evolving MW dSphs and the three largest M31 dSphs along the tidal evolution tracks from \citet{errani2015tidaltracks}. Assuming that none of the dwarfs shown have suffered stripping already, we evolve them according to the tidal evolution tracks, with each dot along the track corresponding to a consecutive 90\% mass loss. Using the same evolution tracks, we infer theoretical progenitors for Crater~II, as well as the corresponding mass loss necessary for such progenitors to reach the velocity dispersion and half-light radius of Crater~II. The shaded region represents the space of possible Crater~II progenitors, color-coded by the remaining mass of the progenitor once it evolves to the position of Crater~II. The gray dash-dotted lines correspond to lines of constant density within the half-light radius. In order of increasing slope, each line corresponds a density of $10^6$~M$_{\odot}$~kpc$^{-3}$, $10^7$~M$_{\odot}$~kpc$^{-3}$, $10^8$~M$_{\odot}$~kpc$^{-3}$, and $10^9$~M$_{\odot}$~kpc$^{-3}$. Data for the MW dSphs are taken from the \citet{munoz2018imaging} catalog, where the half-light radii are derived from fitting Plummer profiles. Data for the three M31 dSphs are taken from \citet{collins2013andromeda}, \citet{tollerud2013splash} and \citet{martin2016LG}. (Left) Results from tidal evolution tracks for cuspy dark matter halos. (Right) Results from tidal evolution tracks for cored dark matter halos.}
\label{fig:craIIevo}
\end{figure*}

\par \citet{penarrubia2008tidalevo}, \citet{errani2015tidaltracks},  \citet{fattahi2018tidalstripping} and \citet{sanders2018crater} showed that tidally stripped dwarf galaxies follow a universal tidal evolution track, where the structural parameters depend only on the fraction of mass lost. \citet{sanders2018crater} in particular confirmed the applicability of tidal evolution tracks for flattened progenitors with cuspy dark matter halos. Applying these tracks to the case of Crater~II, \citet{sanders2018crater} suggested that Crater~II must be tidally stripped to be explainable within the $\Lambda$CDM model. The results from our tidal radius calculation and orbital parameter computation are fully consistent with the hypothesis that Crater~II has experienced tidal interactions with the Milky Way. 

\par Next, we attempt to infer the progenitor of Crater~II by using the tidal evolution tracks of \citet{errani2015tidaltracks}. These tracks were fitted to the tidal evolution of a spherical dark matter halo for both the case of a cored and a cusped progenitor.\footnote{The cuspy tidal evolution tracks from \citet{errani2015tidaltracks} are similar to the \citet{penarrubia2008tidalevo} tracks, but differ from the fitted cuspy tidal evolution tracks from \citet{sanders2018crater}. The chief difference between these two models is that the cusped tidal evolution tracks of \citet{sanders2018crater} never show an expansion of the half-light radius. For the case of flattened, cored halos, the tidal evolution tracks are more difficult to parameterize because they also depend on the inner slope and orbital properties of the satellite (J. Sanders 2018, private communication). Since this exercise cannot promise precise inferences for the progenitors of Crater~II, we use the \citet{errani2015tidaltracks} tracks for the sake of simplicity.} For the evolution of both cored and cuspy progenitors, the half-light radius increases within the first 90\% mass loss. During the process of tidal evolution, the half-light radii of cuspy dark matter halos will increase by up to 25\%, while those of cored dark matter halos can expand by as much as a factor of 4. 

\par First, we attempt to identify analogs to the progenitor of Crater~II from among the currently-known dwarf galaxies in the Local Group. Assuming that none of the MW classical dwarf spheroidal (dSph) galaxies, besides the obvious case of Sagittarius, have experienced significant tidal stripping, we evolve them along the tidal evolution tracks for both cored and cusped progenitors. We perform the same exercise for the three largest dSphs of M31, and present the results for the cusped and the cored cases in the left and right panels of Figure \ref{fig:craIIevo}, respectively. For a cuspy profile, stripping $\sim$90\% of the mass from Andromeda~XXIII or Andromeda~XXI would produce a remnant resembling Crater~II. For a cored dark matter profile, we find that four of the classical dSphs (Leo I, Sculptor, Sextans, and Ursa Minor) may be appropriate progenitor analogs to Crater~II after losing between 70\% and 90\% of their mass. 

\par We then infer where theoretical progenitors of Crater~II would fall in the $r_h$-$\sigma_{Vlos}$ plane. We consider points on an ellipsoidal grid centered on Crater~II at 1066~pc and 2.7~\kms, with major axes of 100~pc and 0.4~\kms. The size of these axes reflect the respective 1 $\sigma$ uncertainty on each of these measurements. The colored swaths in the two panels of Figure \ref{fig:craIIevo} illustrate the result of this analysis, where each prospective progenitor is also color-coded by its remaining fractional mass by the time it becomes an object like Crater~II. Possible cuspy progenitors of Crater~II tend to have lower average densities within their half-light radii than those of the Milky Way dSphs. On the other hand, cored progenitors of Crater~II should resemble Milky Way dSphs in density. 

\subsection{Hercules}
\label{sec:hercorb}

\par
\begin{figure}[!ht] 
\includegraphics[scale=0.5]{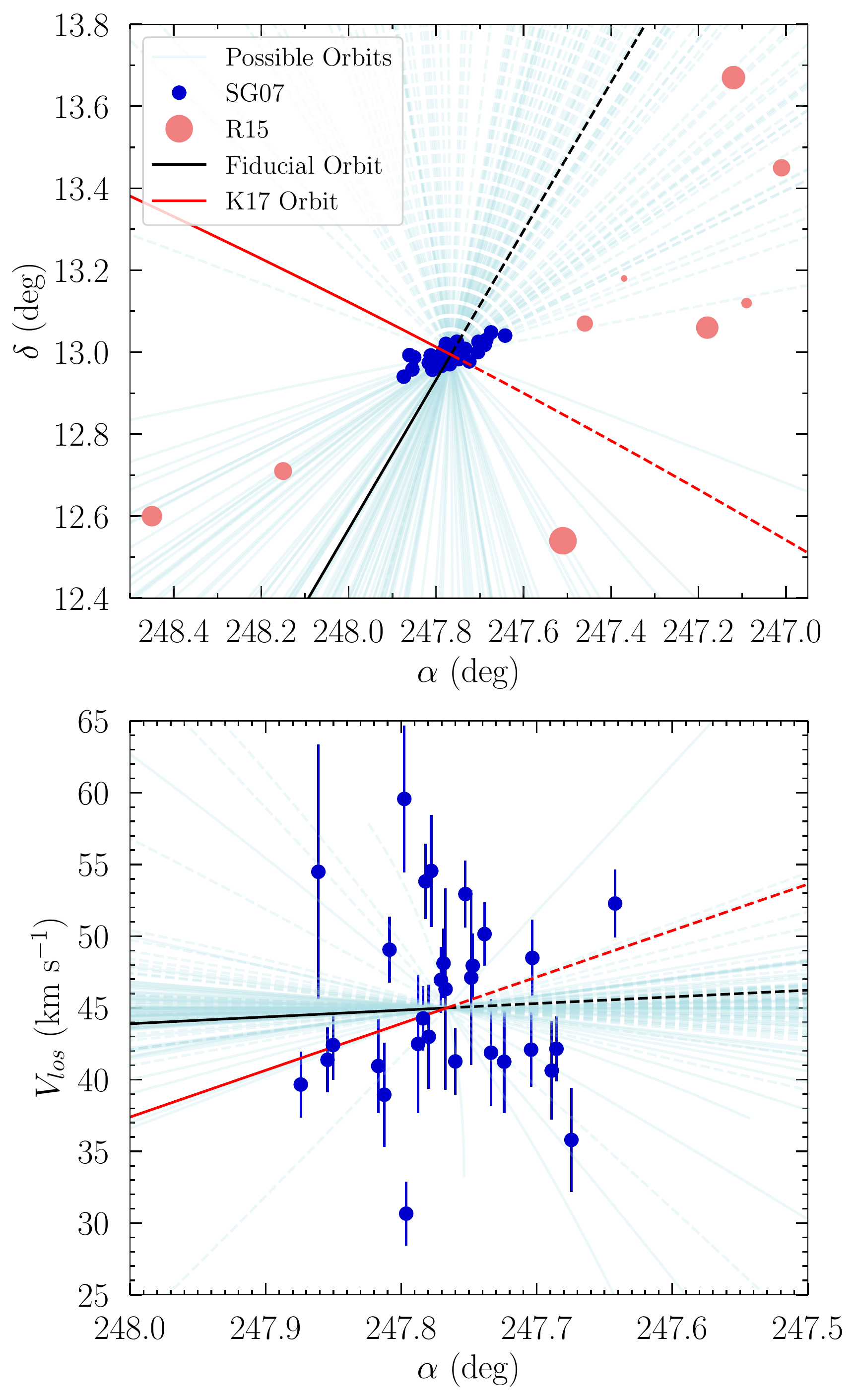}
\caption{(Top) Fiducial orbit of Hercules on the sky. The red dots represent the extra-tidal overdensities detected by \citet{roderick2015structure}, where the size of each dot is linearly scaled by its significance. Our orbit for Hercules is clearly misaligned with that of K17. (Bottom) Line-of-sight velocity of Hercules as a function of RA. The fiducial orbit does not result in a significant velocity gradient across the dwarf, which is consistent with the kinematic data from \citet{simongeha2007}.}
\label{fig:hercobserved}
\end{figure}

 In Figure \ref{fig:hercobserved}, we present the fiducial orbit of Hercules projected on the sky, as well as the line-of-sight velocity of the orbit as a function of position. For reference, we also compare our orbit to the orbit predicted by \citet[][henceforth K17]{kuepper2017orbit} in the case of a tidally-disrupting Hercules. Our orbit is misaligned with the K17 orbit. While our orbit predicts a velocity gradient of 0.6 km~s$^{-1}$~kpc$^{-1}$ across the body of the dwarf, the K17 orbit, with a pericenter distance of 5~kpc, predicts a velocity gradient of 4.9 km~s$^{-1}$~kpc$^{-1}$. Both measurements are inconsistent with the finding from \citet{aden2009herculesmass} of a velocity gradient of 16 $\pm$ 3 km~s$^{-1}$~kpc$^{-1}$. Available spectroscopic data are insufficient for detecting the velocity gradient predicted by either orbit, suggesting that the internal kinematics of Hercules do not currently constrain the possibility of tidal disruption.

\par
\begin{figure*}
\epsscale{1.2}
\plotone{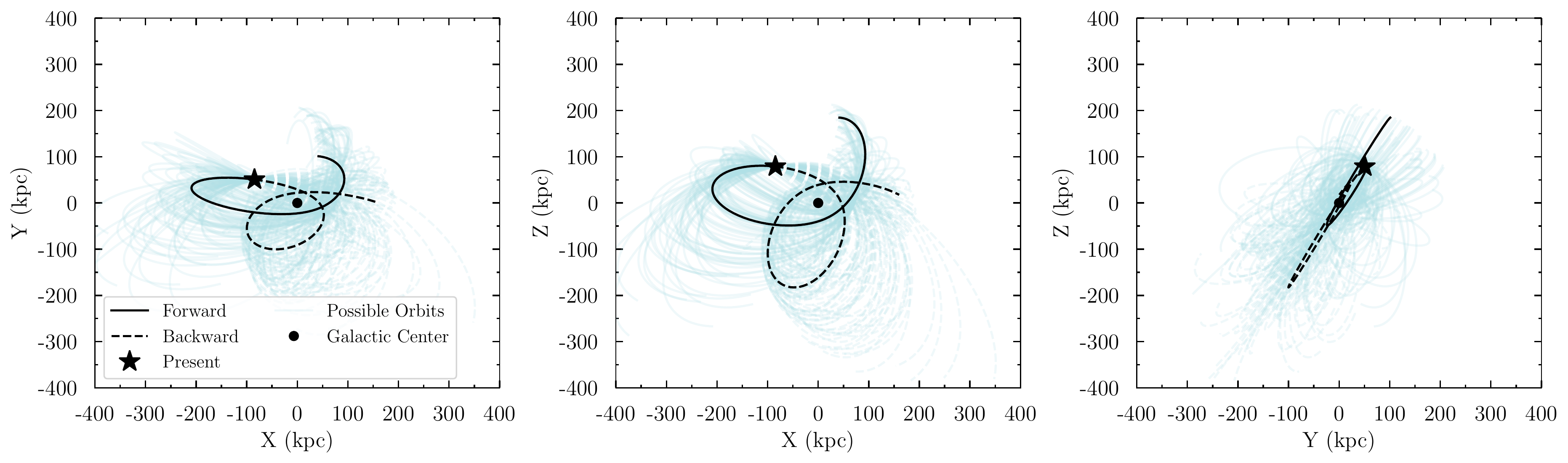}
\caption{Fiducial orbit for Hercules (black), integrated forward and backward for 5~Gyr. In this orbit, Hercules passed pericenter 0.54 Gyr ago, at a distance of 42~kpc from the Milky Way center. Light-blue orbits correspond to other possible orbits drawn from our Monte Carlo simulations sampling the proper motion uncertainties.}
\label{fig:hercfid}
\end{figure*}

\begin{figure*}
\epsscale{1.2}
\plotone{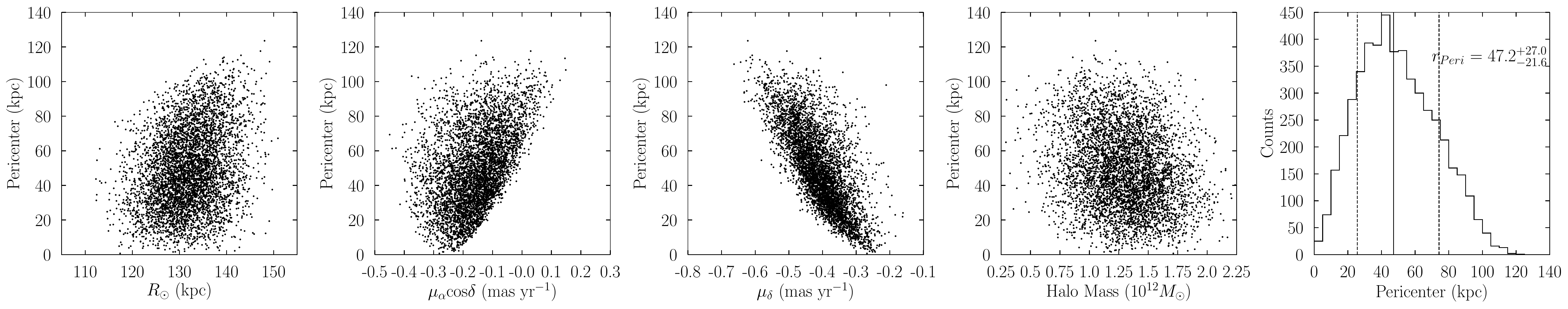}
\caption{Monte Carlo simulations for the orbital pericenter of Hercules, integrated in a spherical potential. The pericenter distance of Hercules's orbit is heavily dependent on $\mu_{\alpha}\cos{\delta}$ and $\mu_{\delta}$: smaller $\mu_{\alpha}\cos{\delta}$ and larger $\mu_{\delta}$ values would result in pericenter distances sufficiently close for tidal disruption. The lines shown on the histogram panel correspond to the 16th, 50th, and 84th percentiles, respectively. 38\% of orbits have pericenter distances smaller than 40~kpc, suggesting that the dwarf may have experienced tidal stripping.}
\label{fig:hercMC}
\end{figure*}

 In Figure \ref{fig:hercfid}, we present the fiducial orbit of Hercules in Galactic coordinates. In this fiducial orbit, Hercules passed pericenter 0.54 Gyr ago, at a distance of 42~kpc from the Milky Way center. In Figure \ref{fig:hercMC}, we present the results of our Monte Carlo simulation for Hercules. Of the 5000 samples, 38\% of the orbits have pericenter distances less than 40~kpc, which suggests that given the Milky Way parameters used in this study, it is possible for Hercules to have been tidally stripped. We note that a more flattened Milky Way halo results in a higher fraction of orbits having pericenters below 40~kpc. 

\par K17 predicted that for Hercules to be tidally disrupted according to their scenario, the dwarf must have a proper motion of $(\mu_{\alpha} \cos{\delta}, \mu_{\delta}) = (-0.21^{+0.019}_{-0.013}$, $-0.24^{+0.015}_{-0.016})$~\masyr. While our point estimates of the proper motion of Hercules are different from their results, particularly in the declination direction, the correlations that emerge from our Monte Carlo simulations are consistent with the findings of K17. That is, proper motions that are increasingly negative in the RA direction and increasingly positive in the Dec direction relative to our point estimates could result in pericenter distances sufficiently small for tidal disruption to occur. 

\subsection{Tidal Evolution of Hercules}
\label{sec:hercevo}

\begin{figure*}
\epsscale{1.2}
\plotone{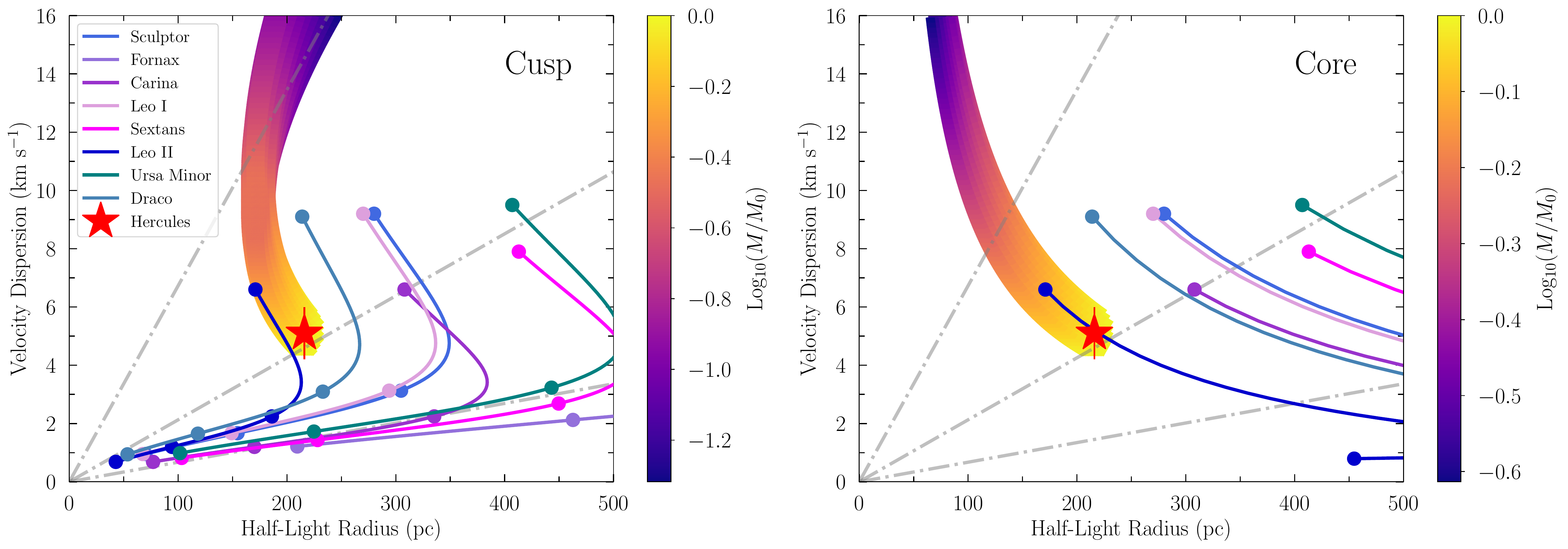}
\caption{Possible tidal evolution of Hercules. We infer possible analogs to the progenitor of Hercules by evolving MW dSphs along the tidal evolution tracks from \citet{errani2015tidaltracks}. Assuming that none of the dwarfs shown have suffered stripping already, we evolve them according to the tidal evolution tracks, with each dot along the track corresponding to a consecutive 90\% mass loss. Using the same evolution tracks, we infer theoretical progenitors for Hercules, as well as the corresponding mass loss necessary for such progenitors to reach the velocity dispersion and half-light radius of Hercules. The shaded region represents the space of possible Hercules progenitors, color-coded by the remaining mass of the progenitor once it evolves to the position of Hercules. The gray dash-dotted lines correspond to lines of constant density within the half-light radius. In order of increasing slope, each line corresponds an average density of $10^7$~M$_{\odot}$~kpc$^{-3}$, $10^8$~M$_{\odot}$~kpc$^{-3}$, and $10^9$~M$_{\odot}$~kpc$^{-3}$. Data for the MW dSphs are taken from the \citet{munoz2018imaging} catalog, where the half-light radii are derived from Plummer profiles. (Left) Results from using tidal evolution tracks for cuspy dark matter halos. (Right) Results from using tidal evolution tracks for cored dark matter halos.}
\label{fig:hercevo}
\end{figure*}

\par The Hercules ultra-faint dwarf (UFD) has long been speculated to be a tidally disrupted object. Our orbital calculations show that it is possible for Hercules to have experienced tidal interactions with the Milky Way. We therefore perform the same tidal evolution track investigation for Hercules that we did for Crater~II, and present our results in Figure \ref{fig:hercevo}. 

\par First, we attempt to identify analogs to the progenitor of Hercules among the currently-known dwarf galaxies. For both the cuspy and cored dark matter profile cases, we find that Leo~II is the closest match to a possible progenitor of Hercules. In the cusped case, an object like Leo~II would have to lose $\sim$70\% of its mass to evolve into an object like Hercules. In the cored case, Hercules lies along the tidal evolution track of Leo~II at the position of 20\% mass loss\footnote{Of course, the stellar mass of Hercules is a factor of 37 smaller than that of Leo~II \citep{munoz2018imaging}, so this hypothetical progenitor would be Leo~II-like in terms of its size and velocity dispersion but with a significantly lower stellar mass.}. 

\par We then infer where theoretical progenitors of Hercules would fall in the $r_h$-$\sigma_{Vlos}$ plane. We consider points on an ellipsoidal grid centered on Hercules at 216~pc and 5.1~\kms, with major axes of 17~pc and 0.9~\kms. The size of these axes reflect the respective 1 $\sigma$ uncertainty on each of these measurements. The half-light radii measurements were adopted from \citet{munoz2018imaging} for the Plummer model, and the velocity dispersion measurement was taken from \citet{simongeha2007}. The colored patches in the two panels of Figure \ref{fig:craIIevo} represent the result of this analysis, where each prospective progenitor is also color-coded by its remaining fractional mass by the time it becomes an object like Hercules. Progenitors corresponding to heavy ($\gtrsim90$\%) mass loss would have to be denser than known Milky Way dwarf spheroidal galaxies. 

\section{Discussion}
\label{sec:discussion}

\subsection{Crater II}
\label{sec:craIIdiscussion}

\par Tidal stripping has been invoked to explain the low velocity dispersion and diffuse, extended size of Crater~II (\citealt{sanders2018crater}, \citealt{fattahi2018tidalstripping}). The results of our orbit analysis suggest that it is very feasible for Crater~II to make pericenter passages sufficiently close to the Galactic Center for tidal stripping to occur. In fact, thanks to its large size, Crater~II may suffer stripping even at its present distance. Although our proper motion measurement for Crater~II is not entirely compatible with the measurement from \citet{kallivayalil2018gaiadr2dsph}, the correlation between pericenter distance and proper motion for Crater~II (Figure \ref{fig:craIIMC}) suggests that all existing proper motion measurements of Crater~II in the literature are consistent with the tidal stripping hypothesis. 

\par Studies of the RR Lyrae population in Crater~II indicate that its stellar populations are similar to those of the Milky Way dSphs (\citealt{joo2018rrlyr}, \citealt{monelli2018rrlyr}). This conclusion is consistent with the simulations that qualitatively recreate observed features of Crater~II by subjecting a dSph-like progenitor to tidal stripping (\citealt{sanders2018crater}, \citealt{fattahi2018tidalstripping}). This conclusion is also consistent with our analysis from Section \ref{sec:craIIevo}, in which we demonstrate using tidal evolution tracks that progenitors of Crater~II that have cored dark matter profiles should resemble Milky Way dSphs in density. 

\citet{rocha2012infall} showed that there is a strong correlation between the binding energy of a subhalo and when it was accreted by its host halo. Using Equation 1 and Figure 1 from that study, we calculate the binding energy for Crater~II and estimate that the galaxy fell into the Milky Way between 4 and 8 Gyr ago. Given the derived orbital period of 2.2 Gyr (Table \ref{tab:craIIorb}), Crater~II may have made multiple pericentric passages around the Milky Way, and experienced several episodes of tidal stripping. This is consistent with the results from Section \ref{sec:craIIevo}, as well as the results of S18, which suggest that the progenitor of Crater~II must have experienced heavy mass loss. 

\par To date, the only known dSph that is conclusively undergoing tidal stripping is Sagittarius (e.g., \citealt{ibata1994}, \citealt{majewski2003}). If Crater~II has undergone similar stripping, as we conclude, then tidal debris associated with Crater~II should be spread along its orbit. Current stellar density maps do not reveal clear evidence of tidal debris, but since Crater~II already falls at the detection limit for existing photometric surveys, the surface brightness of any stellar streams related to Crater~II may be too low to be detected in available imaging data. The recent discovery of the even lower surface brightness dwarf Antlia II \citep{torrealba2018ant2} suggests that a search for stars stripped from Crater~II using proper motion cuts from Gaia could be interesting. 

\subsection{Hercules}
\label{sec:hercdiscussion}

\par Since its discovery, the elongated shape of Hercules has prompted speculation that the UFD has experienced strong tidal interactions with the Milky Way. This hypothesis has inspired photometric follow-up studies in search of extratidal debris (e.g., \citealt{sand2009herculesstructure}, \citealt{roderick2015structure}), as well as a lineage of theoretical studies that attempt to reproduce the observed features of Hercules under the assumption that it has undergone tidal disruption (e.g., \citealt{martin2010herculesdisrupt}, \citealt{kuepper2017orbit}, \citealt{blana2015orb}). 

\par One of the strongest pieces of evidence that would favor the tidal disruption hypothesis is if the orbit of Hercules has a very small pericenter distance, which is characteristic of other ultra-faint dwarf galaxies that are undergoing tidal disruption (e.g., \citealt{simon2018}, \citealt{erkal2018tucIII}, \citealt{carlin2018booIII}). Using the measurements from \citet{simongeha2007} and \citet{munoz2018imaging}, we estimate that an object with the half-light radius and enclosed mass of Hercules would need to approach within at least 40~kpc of the Milky Way center before its stellar component would experience significant tidal effects. The results of our analysis suggest that there is 38\% probability for Hercules's orbit to have a pericenter of less than 40~kpc. Thus, it is plausible, although not certain, that the stellar component of Hercules has been affected by the Galaxy's tidal forces.

\par Other telltale signs of tidal disruption could include a velocity gradient along the stream component (e.g., Tucana III, \citealt{li2018tucIII}). \citet{aden2009herculesmass} claimed to detect a velocity gradient of $16\pm3$~\kms~kpc$^{-1}$ along the body of the dwarf. If the Hercules dwarf contains a stream component, then our fiducial orbit suggests a velocity gradient of 0.6~\kms~kpc$^{-1}$ across the main body of the dwarf. The orbit determined by K17, which has a pericenter distance of 5~kpc, predicts a velocity gradient of 4.9~\kms~kpc$^{-1}$. Both of these velocity gradients are inconsistent with the \citet{aden2009herculesmass} measurement. We show in Figure \ref{fig:hercobserved} that the currently available stellar kinematics for Hercules are not sufficient to detect a velocity gradient of this size. Better velocity data will be needed to assess the kinematics and structure of Hercules.

\par With regard to investigating the dwarf's tidal features, we fail to detect Hercules members in the two largest overdensities identified by \citet{roderick2015structure}. We note that possible orbits for Hercules are roughly aligned with the direction of the most significant extratidal overdensities detected. Whether Hercules members exist beyond the tidal radius of the dwarf is therefore still an open question, but answers to this question could be determined with higher significance by leveraging a combination of deeper photometry and increasingly precise \textit{Gaia} proper motions in future data releases. Deeper spectroscopic studies targeting the bottom of the Hercules RGB will also be fruitful for investigating this issue. 

\par Although we determine that there is a significant chance that Hercules has been tidally stripped by the Milky Way, we note that its elongated shape is not necessarily due to tidal interactions. Hercules does not structurally resemble known systems that are in the process of tidal disruption. For example, although stellar streams extend from the tidally disrupting UFD Tuc~III, the core of the system appears relatively round \citep{mutlupakdil2018}. While there are no theoretical studies to-date of how UFDs evolve through tidal interactions, the results from \citet{penarrubia2008tidalevo} for classical dSphs suggest that the shape of a galaxy undergoing tidal interactions is preserved until the final stages of disruption. It is therefore possible that the present-day structural features of Hercules trace back to its natal shape. 

\par Finally, we consider what the orbital properties of Hercules imply about its infall history. Using the relationships from \citet{rocha2012infall}, we infer that Hercules fell into the Milky Way $\sim2-4$~Gyr ago. Given its orbital period of 3.5~Gyr (see Table \ref{tab:hercorb}), Hercules has likely made only one pericentric passage around the Milky Way. Increasingly detailed proper motions from \textit{Gaia} in the future will provide stronger constraints on the orbital pericenter of Hercules, which in turn should improve estimates of how much mass it could have lost via stripping. 

\section{Conclusion}
\label{sec:conclusion}

\par In this study, we present Magellan/IMACS spectroscopy of 37 stars in Crater~II, including 12 newly-identified members, within 15$\arcmin$ from the center of the dwarf.  We measure a velocity dispersion of $\sigma_{Vlos} = \craveldisp^{+\craveldispu}_{-\craveldispl}$~\kms, which corresponds to $M/L$ = $\craml^{+\cramlupper}_{-\cramllower}$~M$_{\odot}/$L$_{\odot}$ within its half-light radius. Thus we confirm that Crater~II resides in a kinematically cold dark matter halo. We also attempt to identify stars associated with the Hercules ultra-faint dwarf galaxy within two of the extratidal stellar overdensities detected in previous studies, but fail to confirm any such stars. 

\par Combining member samples from our spectroscopy and the literature, we measure the bulk proper motion of each dwarf galaxy using \textit{Gaia} DR2 astrometry. With the complete 6D phase space information of each dwarf, we test the hypotheses that they have experienced tidal interactions with the Milky Way by investigating whether they could have made sufficiently close perigalacticon approaches. 

\par We find that the perigalacticon distance for Crater~II suggests that it is likely to have been tidally stripped; all currently existing proper motion measurements in the literature are consistent with this result as well. Using tidal evolution tracks, we infer possible analogs to the progenitor of Crater~II. We find that if the progenitor of Crater~II resided in a cuspy dark matter halo, then two M31 satellites, Andromeda~XXIII and Andromeda~XXI, could tidally evolve into an object similar to Crater~II. If the progenitor of Crater~II resided in a cored dark matter halo, then four of the classical MW dSphs, Sculptor, Leo~I, Sextans, and Ursa Minor, may be appropriate analogs to the progenitor of Crater~II. 

\par Since Crater~II has an orbital period of 2.2~Gyr and fell into the Milky Way over 4 Gyr ago, we suggest that like Sagittarius, it may have made multiple pericentric passages around the Milky Way. Follow-up wide-field photometric studies of Crater~II in the era of LSST and WFIRST are promising avenues for revealing stellar streams and providing valuable observational constraints on the tidal stripping of the dwarf. 

\par We find that the perigalacticon distance for Hercules also suggests that the dwarf may have been tidally stripped. Assuming this is true, we also use tidal evolution tracks to infer possible progenitors of Hercules. We find that the dSph Leo~II is an appropriate analog to the progenitor of Hercules for both the cored and cuspy cases. In addition to further photometric, spectroscopic and astrometric data for investigating the tidal history of Hercules, theoretical studies looking at the tidal evolution of UFDs would be useful for providing additional context to the observed morphological diversity of dwarf galaxies.  

\acknowledgements{S.W.F. thanks Gwen Rudie and Jorge Moreno for their supervision throughout this work. She also thanks Jo Bovy, Alex Ji, Andrew Benson, Jason Sanders, Manoj Kaplinghat, and Jorge Pe{\~n}arrubia for useful conversations regarding the analysis. Finally, she would like to acknowledge the cleaning staff at academic and telescope facilities, especially those whose communities are often excluded from academic pursuits, but whose labor maintains the spaces where astrophysical inquiry can flourish.

Funding for S.W.F. was provided by the Rose Hills Foundation. J.D.S. acknowledges support from the National Science Foundation through grant AST-1714873. A.G.A acknowledges financial support from Carnegie Observatories through the Carnegie-Chile fellowship.

\par This work has made use of data from the European Space Agency (ESA) mission
{\it Gaia} (\url{https://www.cosmos.esa.int/gaia}), processed by the {\it Gaia}
Data Processing and Analysis Consortium (DPAC,
\url{https://www.cosmos.esa.int/web/gaia/dpac/consortium}). Funding for the DPAC
has been provided by national institutions, in particular the institutions
participating in the {\it Gaia} Multilateral Agreement.

\par The Pan-STARRS1 Surveys (PS1) and the PS1 public science archive have been made possible through contributions by the Institute for Astronomy, the University of Hawaii, the Pan-STARRS Project Office, the Max-Planck Society and its participating institutes, the Max Planck Institute for Astronomy, Heidelberg and the Max Planck Institute for Extraterrestrial Physics, Garching, The Johns Hopkins University, Durham University, the University of Edinburgh, the Queen's University Belfast, the Harvard-Smithsonian Center for Astrophysics, the Las Cumbres Observatory Global Telescope Network Incorporated, the National Central University of Taiwan, the Space Telescope Science Institute, the National Aeronautics and Space Administration under Grant No. NNX08AR22G issued through the Planetary Science Division of the NASA Science Mission Directorate, the National Science Foundation Grant No. AST-1238877, the University of Maryland, Eotvos Lorand University (ELTE), the Los Alamos National Laboratory, and the Gordon and Betty Moore Foundation. The authors wish to recognize and acknowledge the very significant cultural role and reverence that the summit of Maunakea has always had within the indigenous Hawaiian community. We are most fortunate to have the opportunity to utilize observations taken from this mountain. 

\par Funding for SDSS-III has been provided by the Alfred P. Sloan Foundation, the Participating Institutions, the National Science Foundation, and the U.S. Department of Energy Office of Science. The SDSS-III web site is http://www.sdss3.org/.

\par SDSS-III is managed by the Astrophysical Research Consortium for the Participating Institutions of the SDSS-III Collaboration including the University of Arizona, the Brazilian Participation Group, Brookhaven National Laboratory, Carnegie Mellon University, University of Florida, the French Participation Group, the German Participation Group, Harvard University, the Instituto de Astrofisica de Canarias, the Michigan State/Notre Dame/JINA Participation Group, Johns Hopkins University, Lawrence Berkeley National Laboratory, Max Planck Institute for Astrophysics, Max Planck Institute for Extraterrestrial Physics, New Mexico State University, New York University, Ohio State University, Pennsylvania State University, University of Portsmouth, Princeton University, the Spanish Participation Group, University of Tokyo, University of Utah, Vanderbilt University, University of Virginia, University of Washington, and Yale University.

\par This research has made use of NASA's Astrophysics Data System Bibliographic Services and the VizieR catalogue access tool, CDS, Strasbourg, France. The original description of the VizieR service was published in A\&AS 143, 23 \citep{vizier}. 

\facilities{Magellan (Baade): IMACS, Gaia}

\software{astropy \citep{astropy}, matplotlib \citep{matplotlib}, pandas \citep{pandas}, scipy \citep{scipy}, emcee \citep{emcee}, COSMOS \citep{cosmos}, galpy \citep{bovy2015galpy}, numpy \citep{numpy}}}

\bibliography{craII,generaldsph,herc,tools}
\bibliographystyle{aasjournal}



\startlongtable
\begin{deluxetable*}{lccccccccccccc}          
\tablecaption{Velocity and Metallicity Measurements for Confirmed Cra II Members}
\tabletypesize{\footnotesize}
\tablehead{\colhead{PS1 objID} & \colhead{MJD} & \colhead{R.A.}  & \colhead{Decl.} & \colhead{$g_{P1}$} & \colhead{$r_{P1}$} & \colhead{S/N} & \colhead{$v_{helio}$}    & \colhead{EW}        & \colhead{[Fe/H]} \\
           \colhead{}          & \colhead{}    & \colhead{(deg)} & \colhead{(deg)} & \colhead{(mag)}      & \colhead{(mag)}      & \colhead{(pixel$^{-1}$)}    & \colhead{(\kms)} & \colhead{(\AA)}       & \colhead{} }
\startdata
PSO J114814.08$-$182502.8 & 58202.3 & 177.05862 & $-$18.41747 & 19.20 & 18.45 &       35.8 &   87.2   $\pm$     1.0 & 4.6 $\pm$ 0.2 & $-$1.54 $\pm$ 0.16 \\
PSO J114817.18$-$182426.9 & 58202.3 & 177.07153 & $-$18.40750 & 19.96 & 19.30 &       20.0 &   88.0   $\pm$     1.2 & 3.8 $\pm$ 0.3 & $-$1.87 $\pm$ 0.20 \\
PSO J114817.66$-$182217.3 & 58202.3 & 177.07355 & $-$18.37151 & 20.79 & 20.29 &        8.8 &   87.6   $\pm$     1.7 & 3.5 $\pm$ 0.5 & $-$2.01 $\pm$ 0.29 \\
PSO J114818.65$-$182754.7 & 58202.3 & 177.07765 & $-$18.46524 & 19.77 & 19.10 &       23.7 &   94.6   $\pm$     1.1 & 3.5 $\pm$ 0.2 & $-$1.99 $\pm$ 0.19 \\
PSO J114820.50$-$183233.3 & 58202.3 & 177.08540 & $-$18.54264 & 20.33 & 19.73 &       13.3 &   86.0   $\pm$     1.3 & 3.5 $\pm$ 0.4 & $-$2.00 $\pm$ 0.22 \\
PSO J114821.07$-$183604.0 & 58202.3 & 177.08775 & $-$18.60114 & 19.75 & 19.07 &       21.6 &   86.1   $\pm$     1.1 &           ... &              ... \\
PSO J114822.04$-$183028.8 & 58202.3 & 177.09180 & $-$18.50804 & 20.75 & 20.21 &        9.8 &   88.2   $\pm$     1.6 & 4.3 $\pm$ 0.6 & $-$1.67 $\pm$ 0.32 \\
PSO J114824.74$-$182208.5 & 58202.3 & 177.10307 & $-$18.36906 & 19.03 & 18.27 &       40.8 &   92.9   $\pm$     1.0 & 3.5 $\pm$ 0.1 & $-$2.00 $\pm$ 0.16 \\
PSO J114824.84$-$182642.3 & 58202.3 & 177.10345 & $-$18.44512 & 21.16 & 20.70 &        5.2 &   87.2   $\pm$     2.8 &           ... &              ... \\
PSO J114825.96$-$183223.5 & 58202.3 & 177.10812 & $-$18.53990 & 20.04 & 19.38 &       18.6 &   87.3   $\pm$     1.2 & 3.9 $\pm$ 0.2 & $-$1.82 $\pm$ 0.18 \\
PSO J114828.75$-$182345.1 & 58202.3 & 177.11977 & $-$18.39589 & 21.12 & 20.48 &        6.8 &   89.3   $\pm$     1.8 & 3.7 $\pm$ 0.5 & $-$1.91 $\pm$ 0.27 \\
PSO J114829.71$-$182328.3 & 58202.3 & 177.12377 & $-$18.39123 & 19.98 & 19.35 &       18.4 &   87.5   $\pm$     1.2 & 4.0 $\pm$ 0.3 & $-$1.78 $\pm$ 0.20 \\
PSO J114829.90$-$182402.2 & 58202.3 & 177.12456 & $-$18.40066 & 19.85 & 19.20 &       19.8 &   90.9   $\pm$     1.1 & 3.2 $\pm$ 0.3 & $-$2.13 $\pm$ 0.20 \\
PSO J114830.71$-$182912.3 & 58202.3 & 177.12795 & $-$18.48681 & 20.94 & 20.33 &        6.8 &   90.3   $\pm$     1.6 &           ... &              ... \\
PSO J114837.99$-$182654.6 & 58202.3 & 177.15827 & $-$18.44855 & 21.05 & 20.74 &        5.2 &   82.4   $\pm$     3.6 &           ... &              ... \\
PSO J114850.25$-$182956.9 & 58203.3 & 177.20932 & $-$18.49917 & 19.91 & 19.30 &       16.9 &   89.3   $\pm$     1.2 & 2.9 $\pm$ 0.2 & $-$2.29 $\pm$ 0.19 \\
PSO J114902.20$-$182832.2 & 58203.3 & 177.25915 & $-$18.47564 & 20.99 & 20.50 &        6.2 &   83.6   $\pm$     2.5 & 2.4 $\pm$ 1.0 & $-$2.53 $\pm$ 0.53 \\
PSO J114905.30$-$182912.7 & 58203.3 & 177.27205 & $-$18.48689 & 19.41 & 18.71 &       27.5 &   85.2   $\pm$     1.0 & 3.9 $\pm$ 0.2 & $-$1.81 $\pm$ 0.17 \\
PSO J114906.67$-$182936.3 & 58203.3 & 177.27777 & $-$18.49345 & 20.77 & 20.17 &        8.3 &   87.3   $\pm$     1.8 & 2.8 $\pm$ 0.6 & $-$2.31 $\pm$ 0.32 \\
PSO J114915.29$-$183808.0 & 58203.3 & 177.31371 & $-$18.63558 & 20.38 & 19.92 &        9.6 &   86.8   $\pm$     1.9 & 2.6 $\pm$ 0.2 & $-$2.44 $\pm$ 0.19 \\
PSO J114916.21$-$183228.7 & 58203.3 & 177.31752 & $-$18.54131 & 21.06 & 20.67 &        5.2 &   90.8   $\pm$     2.2 &           ... &              ... \\
PSO J114917.07$-$181413.5 & 58203.3 & 177.32109 & $-$18.23711 & 19.50 & 18.81 &       12.1 &   88.5   $\pm$     1.3 & 3.6 $\pm$ 0.4 & $-$1.94 $\pm$ 0.22 \\
PSO J114917.45$-$183756.9 & 58203.3 & 177.32270 & $-$18.63248 & 19.41 & 18.72 &       24.4 &   92.2   $\pm$     1.1 & 4.1 $\pm$ 0.3 & $-$1.75 $\pm$ 0.19 \\
PSO J114919.00$-$181145.4 & 58203.3 & 177.32913 & $-$18.19598 & 19.45 & 18.80 &       10.3 &   84.8   $\pm$     1.3 & 4.8 $\pm$ 0.4 & $-$1.43 $\pm$ 0.23 \\
PSO J114919.17$-$181658.4 & 58203.3 & 177.32982 & $-$18.28291 & 19.09 & 18.35 &       17.6 &   83.5   $\pm$     1.1 & 4.7 $\pm$ 0.3 & $-$1.49 $\pm$ 0.19 \\
PSO J114922.14$-$183048.4 & 58201.3 & 177.34222 & $-$18.51345 & 19.65 & 19.03 &       16.6 &   90.2   $\pm$     1.1 & 3.1 $\pm$ 0.3 & $-$2.17 $\pm$ 0.20 \\
PSO J114922.24$-$183225.9 & 58201.3 & 177.34267 & $-$18.54054 & 18.70 & 17.86 &       41.1 &   89.0   $\pm$     1.0 & 3.5 $\pm$ 0.1 & $-$2.01 $\pm$ 0.17 \\
PSO J114924.62$-$183732.5 & 58203.3 & 177.35259 & $-$18.62571 & 20.33 & 19.78 &       10.3 &   88.6   $\pm$     2.1 & 2.4 $\pm$ 0.6 & $-$2.55 $\pm$ 0.36 \\
PSO J114924.77$-$183139.2 & 58201.3 & 177.35319 & $-$18.52755 & 20.62 & 20.05 &        6.0 &   83.7   $\pm$     2.0 & 3.0 $\pm$ 0.8 & $-$2.24 $\pm$ 0.41 \\
PSO J114927.13$-$183415.5 & 58203.3 & 177.36303 & $-$18.57098 & 21.20 & 20.67 &        5.4 &   84.9   $\pm$     2.0 & 2.9 $\pm$ 0.7 & $-$2.29 $\pm$ 0.38 \\
PSO J114928.96$-$182939.3 & 58201.3 & 177.37065 & $-$18.49424 & 20.12 & 19.62 &       10.2 &   83.9   $\pm$     1.5 & 3.2 $\pm$ 0.3 & $-$2.12 $\pm$ 0.20 \\
PSO J114938.64$-$183003.0 & 58201.3 & 177.41098 & $-$18.50083 & 20.62 & 20.15 &        5.9 &   89.5   $\pm$     3.1 &           ... &              ... \\
PSO J114941.89$-$182843.1 & 58201.3 & 177.42453 & $-$18.47863 & 19.86 & 19.20 &       15.6 &   83.0   $\pm$     1.2 & 3.6 $\pm$ 0.3 & $-$1.97 $\pm$ 0.22 \\
PSO J114950.38$-$182359.6 & 58201.3 & 177.45992 & $-$18.39988 & 18.70 & 17.86 &       42.2 &   86.0   $\pm$     1.0 & 4.3 $\pm$ 0.1 & $-$1.63 $\pm$ 0.16 \\
PSO J114955.88$-$182356.9 & 58201.3 & 177.48285 & $-$18.39914 & 20.10 & 19.52 &       11.6 &   85.4   $\pm$     1.8 &           ... &              ... \\
PSO J114959.68$-$182745.1 & 58201.3 & 177.49867 & $-$18.46251 & 19.39 & 18.75 &       21.6 &   82.5   $\pm$     1.2 & 3.4 $\pm$ 0.6 & $-$2.04 $\pm$ 0.31 \\
PSO J115010.24$-$182825.8 & 58201.3 & 177.54268 & $-$18.47382 & 19.92 & 19.36 &       13.9 &   86.0   $\pm$     1.3 & 2.3 $\pm$ 0.5 & $-$2.58 $\pm$ 0.31 \\
\label{tab:cra2mem}
\enddata
\tablecomments{For easier comparison to existing catalogs, the PS1 photometry in this table has not been corrected for extinction. PSO J114820.50-183233.3, PSO J114825.96-183223.5, and PSO J114829.90-182402.2 are possible binary star candidates because our velocity measurements for these stars differ from those of C17 by $\gtrsim2 \sigma$.}
\end{deluxetable*}

\end{document}